\documentclass[a4paper]{article}
\usepackage[psamsfonts]{amssymb}
\usepackage{amsmath}
\author{H. Mohseni Sadjadi\footnote{hmohseni@phymail.ut.ac.ir} and M.Alimohammadi\footnote{alimohmd@ut.ac.ir}
\\ {\small Department of Physics, University of Tehran,}
\\ {\small North Karegar Ave., Tehran, Iran.}}
\title{ Electrostatic self-energy and Bekenstein entropy bound in the massive
Schwinger model}
\begin{document}
\maketitle
\begin{abstract}
We obtain the electrostatic energy of two opposite charges near
the horizon of stationary black-holes in the massive Schwinger
model. Besides the confining aspects of the model, we discuss the
Bekenstein entropy upper bound of a charged object using the
generalized second law. We show that despite the massless case, in
the massive Schwinger model the entropy of the black hole and
consequently the Bekenstein bound are altered by the vacuum
polarization.
\end{abstract}
\section{Introduction}
Using the generalized second law for black holes, a universal
upper bound on the entropy of a macroscopic charged object can be
obtained by a gedanken experiment known as Geroch process. In this
process the object is lowered adiabatically toward the horizon and
then is assimilated from a small proper distance into the hole
\cite{bek0},\cite{bek1}. On the basis of the no-hair theorem it is
claimed that this bound cannot be improved by other quantum
numbers which are carried by the object, e.g. the baryon number
\cite{bek2}. A direct proof for the special case of a particle
with a scalar charge, is given in \cite{mayo}. This proof is based
on the fact that the self-energy receives no contribution from the
scalar charge and the entropy of the assimilated object consists
only of its gravitational energy.

A quantum effect relevant to this subject is the radiative
correction involving the loop graphs, the so called vacuum
polarization. Because of computational difficulties in studying
the quantum effects in four dimensional curved space-times, one
can consider lower dimensional models as a framework to explore
these effects and to obtain the physical clues to the real four
dimensional cases.

In \cite{moh} it is shown that in the massless Schwinger model
\cite{sch}, the Bekenstein entropy upper bound is not affected by
vacuum polarization. This lies on the fact that the effect of
vacuum polarization is only appeared in the mass gained by the
gauge field. This can not change the entropy upper bound since the
gauge field behaves like a conformal massless field near the
horizon.

In this paper, using the generalized second law for black-holes,
we compute the entropy upper bound of a charged macroscopic object
in the massive Schwinger model \cite{col}-\cite{sad2}.

The paper is organized as follows: The section two, is devoted to
a brief introduction of massive Schwinger model in curved
space-time. In the section three we discuss the confining and
screening aspects of this model in static black-hole background.
We show that our results asymptotically reduce to ones obtained in
\cite{ali} where the WKB approximation has been used. Also we
compute the electrostatic self-energy of a charged particle very
close to the horizon in terms of Bessel functions. These results
are used to check the more complicated expression of self-energy
derived in the section four where a term proportional to scalar
curvature is included in the metric expansion. Our method is based
on Taylor expansion of the metric in Schwarzschild gauge near the
horizon, leading to Rindler (section three) or AdS space-time
(section four) depending on the order of our approximation.
Finally in the section five, we find out when an external charge
is assimilated into the hole, the massive dynamical fermions
affects the hole's entropy via the self-energy of the charge. This
effect disappears when dynamical fermions are massless.

In this paper the charges of the hole and the object are assumed
to be very small in the scale of the hole mass.  The units $c= G =
\hbar =1$ are used throughout the paper.
\section{preliminaries}
Since all two dimensional spaces are conformally flat, the most
general stationary metric reads
\begin{equation} \label{1}
ds^2=\sqrt{g(x)}\left(dt^2-dx^2 \right),
\end{equation}
where $g(x)=|{\rm det}g_{\mu\nu}(x)|$, and $g_{\mu \nu}$   depends
only on the spatial coordinate $x$. On this space-time, the
quantum electrodynamics of fermions of mass $m$ and charge $e$ is
described by the action \cite{ebo}, \cite{ali}
\begin{equation}\label{2}
S=\int \left[-\frac{1}{4}g^{\mu \nu}g^{\lambda \beta} F_{\mu
\lambda}F_{\nu
\beta}+i\bar{\psi}\gamma^\mu\left(\nabla_\mu-ieA_\mu\right) \psi -
m\bar{\psi}\psi\right]\sqrt{g(x)}dxdt.
\end{equation}
$\gamma^\mu=e^{\mu a}\gamma_a$ is the curved space-time
counterparts of Dirac gamma matrices $\gamma_a$ and the zweibeins
$e^{\mu a}$'s satisfy $e^{\mu a}e_{\nu a}=\delta_\nu^\mu$. The
covariant derivative, acting on fermions, is given by
$\nabla_\mu=\partial_\mu+{1\over 2}\omega_{\mu}^{ab}\sigma_{ab}$
in which $\sigma_{ab}={1\over 4}[\gamma_a,\gamma_b]$ and
$\omega_\mu^{ab}$'s are spin-connections. $F_{\mu
\nu}=\partial_\mu A_\nu-\partial_\nu A_\mu$ are the field strength
tensor components. The vacuum angle is assumed to be zero.

The corresponding bosonic action, in the presence of covariant
conserved current
\begin{equation} \label{3}
J^\mu
=\frac{e'}{\sqrt{g(x)}}\left[\delta(x-a)-\delta(x-b)\right]\delta^\mu_0,
\end{equation}
describing two opposite external charges $e'$ and $-e'$ located at
$x=a$ and $x=b$, respectively, is \cite{ali}
\begin{eqnarray} \label{4}
S&=&\int[\frac{1}{2}\sqrt{g}g^{\mu \nu}\partial_\mu \phi
\partial_\nu \phi
+\frac{e}{\sqrt{\pi}}F\phi+\frac{m}{\pi}g^{\frac{1}{4}}(x)\exp[-2\pi
G({\bf x},{\bf x})]\nonumber\\
&& N_\mu \cos(2\sqrt{\pi}\phi)+
\frac{1}{2\sqrt{g(x)}}F^2+\frac{e}{\sqrt{\pi}}\eta F ]dxdt.
\end{eqnarray}
The dual field strength $F$ is defined through $F={\hat
{\epsilon}}^{\mu \nu}\partial_\mu A_\nu$, where ${\hat
{\epsilon}}^{01}=-{\hat {\epsilon}}^{10}=1$. $N_{\mu}$ is normal
ordering with respect to scale
\begin{equation} \label{5}
\mu=\frac{e}{\sqrt{\pi}},
\end{equation}
and $G({\bf x},{\bf x}):=G_\mu ({\bf x},{\bf x})- D({\bf x},{\bf
x})$, ${\bf x}=(t,x)$. $G_{\mu}({\bf x},{\bf x})$ is the Green
function of a massive scalar field of mass $\mu$ and
\begin{displaymath}D({\bf x}_i,{\bf x}_j)=-\frac{1}{4\pi}\ln{|{\bf x}_i-{\bf x}_j|}^2.\end{displaymath}
In terms of step functions, $\eta$ is defined as
\begin{displaymath}\eta=\frac{\sqrt{\pi}e'}{e}\left[\theta
(x-b)-\theta(x-a)\right].\end{displaymath} In the Schwarzschild
gauge, the metric
\begin{equation}\label{6}
 ds^2=f(r)dt^2-{\frac{1}{f(r)}}dr^2, \qquad r>h
\end{equation}
describes a black hole whose the horizon is located at $r=h$. $h$
is defined through $f(h)=0$. We assume that $f(r)$ is  a positive
${\mathcal{C}}^\infty$ function for $r>h$. In tortoise coordinates
$(t,x)$, defined by  $ dx^2=dr^2/f^2(r)$, the metric (\ref{6})
reduces to (\ref{1}) for $f(r)=\sqrt{g(x)}$. By integrating over
the gauge fields of action ({\ref{4}), we obtain
\begin{equation}\label{7}
S_{{\rm eff.} }=\int\left[\frac{1}{2}g^{\mu \nu}\partial_{\mu}\phi
\partial_{\nu}\phi+m\Sigma N_{\mu}
\cos(2\sqrt{\pi}\phi)-\frac{\mu^2}{2}(\phi+\eta)^2\right]drdt,
\end{equation}
where $g_{tt}=-1/g_{rr}=f(r)$ and $\Sigma:=\exp[-2\pi G({\bf
x},{\bf x})]/(\pi f^{\frac{1}{2}}(r))$ is the chiral condensate
$<\bar{\psi}\psi>$ \cite{ali}.

The electrostatic energy of external charges can be calculated by
either performing a typical Wilson loop calculation, or by
computing the ground state expectation value of the Hamiltonian in
the presence of external source
\begin{equation} \label{8}
E_{{\rm elec.}}:=<\Omega_Q|H_Q|\Omega_Q>-<\Omega_0|H_0|\Omega_0>.
\end{equation}
$H_0(H_Q)$ and $|\Omega_0>(|\Omega_Q>)$ are the Hamiltonian and
the ground state in the absence (presence) of the test charges,
respectively \cite{abd}. In the static case, the energy of the
system measured by an observer whose two-velocity is parallel to
the global time-like Killing vector of the space-time, i.e.
$u=(f^{-\frac{1}{2}},0)$, is
\begin{equation}\label{9}
E=\int T_0^0 dr=-\int L dr.
\end{equation}
$T_\mu^\nu$ is the energy-momentum tensor and $L$ is the
Lagrangian density. For small $\phi$, where
$\cos(2\sqrt{\pi}\phi)\simeq 1-2\pi \phi^2$, the action (\ref{4})
becomes quadratic with respect to the gauge fields (this can be
seen by integrating out the fermionic degrees of freedom in
(\ref{2}) or the bosonic ones in (\ref{4})). Hence in the static
case, the electrostatic energy of charges in terms of the gauge
field Green function $\mathcal{G}(r,r')$ becomes \cite{sad2}
\begin{equation} \label{10}
E_{{\rm
elec.}}=-\frac{e'^2}{2}\left[\mathcal{G}(r_a,r_a)+\mathcal{G}(r_b,r_b)-
2\mathcal{G}(r_a,r_b)\right].
\end{equation}
$e'^2\mathcal{G}(r_a,r_b)$ is the interaction energy, $E_{{\rm
int.}}$, of two opposite charges $e'$ and $-e'$ located at
$r_a:=r(a)$ and $r_b:=r(b)$, and
$-\left(e'^2/2\right)\mathcal{G}(r,r)$ is the self-energy $E_{{\rm
self.}}$. In this paper, by self-energy of a charge we mean the
change of energy of the system when the charge is added to it.
Note that in contrast to higher dimensional cases,
$\mathcal{G}(x,x)$, where $x$ is the spatial coordinate of the two
dimensional space-time, is not infinite. For example in the
massless Schwinger model on flat space-time, we have
$\mathcal{G}(x,y)=-1/\left(2\mu\right)\exp[-\mu|x-y|]$, which
becomes the constant $\mathcal{G}(x,x)=-1/\left(2\mu\right)$ in
the coincident limit \cite{jac}.

Up to the first order of $m$ and when $m\Sigma\ll \mu^2$, the
electrostatic energy of widely separated charges, i.e.
$\eta=(e'/e)\sqrt{\pi}$, is \cite{col}, \cite{ali}
\begin{equation}\label{11}
E_{{\rm
elec.}}=E(\eta)-E(0)=m\left[1-\cos\left(2\pi\frac{e'}{e}\right)\right]
\int_{r_a}^{r_b}\Sigma dr.
\end{equation}
Note that the action is not quadratic here and the equation
(\ref{10}) is not valid. For constant values of $\eta$, using
(\ref{4}), one can show that the gauge fields with zero winding
number ( $\int Fdrdt=0$) do not contribute to the electrostatic
energy of external charges.

For finitely separated charges, $\eta$ is no more a constant and
the solution of the classical equation of motion for the field
$|\phi|\ll 1$ is
\begin{equation}\label{12}
\phi=-\frac{\mu^2}{\partial_r g^{rr}\partial_r+\mu^2+4\pi m
\Sigma}\eta.
\end{equation}
In this regime, the action becomes Gaussian and the classical
solutions of the action coincide with the quantum ones. Putting
(\ref{12}) back into Lagrangian (\ref{7}), eq.(\ref{9}) results
\begin{equation}\label{13}
E_{{\rm
elec.}}=\frac{\mu^2}{2}\left[\int_{r_a}^{r_b}\left(\eta^2-\mu^2\eta
\frac{1}{\partial_r g^{rr}\partial_r+\mu^2+4\pi m
\Sigma}\eta\right)dr\right].
\end{equation}
This equation can be rewritten as
\begin{equation}\label{14}
E_{{\rm
elec.}}=\left(\frac{e'}{\mu}\right)^2\left[\frac{\mu^2}{2}(r_b-r_a)+
\Pi\right],
\end{equation}
where
\begin{eqnarray}\label{15}
\Pi&:=&-\frac{\mu^4}{2}\int_{r_a}^{r_b}dr\Bigg[\int_{r'=r_a}^{r'=r}G(r_{>}=
r,r_{<}=r')dr'\nonumber \\
&+&\int_{r'=r}^{r'=r_b} G(r_{>}=r',r_{<}=r)dr'\Bigg],
\end{eqnarray}
and
\begin{equation}\label{16}
\left(\partial_rg^{rr}\partial_r+\mu^2+4\pi m\Sigma
\right)G(r,r')=\delta(r,r').
\end{equation}
We denote the smaller (bigger) argument of the Green function by
$>$ ($<$).

In \cite{ali}, for a slowly varying metric
$df^{\frac{1}{2}}(r)/dr\ll {\bar \mu}$, where
\begin{equation}\label{17}
\bar{\mu}^2=\mu^2+4\pi m \Sigma ,
\end{equation}
$E_{{\rm elec.}}$ has been derived as
\begin{eqnarray}\label{18}
E_{{\rm elec.}}&=&\frac{e'^2}{2}\left(1-\frac{\mu^2}{\bar{\mu}^2}
\right)\left(r_b-r_a\right)
+\frac{e'^2\mu^2}{4{\bar{\mu}^3}}\Big[f^{\frac{1}{2}}
\left(r_a\right)\nonumber \\
&+&f^{1\over
2}\left(r_b\right)-2f^{\frac{1}{4}}\left(r_a\right)f^{\frac{1}{4}}
\left(r_b\right)
\exp[-\int^{r_b}_{r_a}\bar{\mu}f^{-\frac{1}{2}}(u)du]\Big].
\end{eqnarray}
The above equation has been obtained in the zeroth order of WKB
approximation . In this small curvature limit, (\ref{18}) is the
leading term of the energy.  If the metric were a constant, the
self-force (the derivative of the self-energy) would vanish and
the energy should become the same as the flat case, shifted by the
metric factor $f^{1/2}$. This result is consistent with the
measurement of a constantly accelerated observer. The condition of
validity of (\ref{18}) is $e'\ll e$ and the first expression in
(\ref{18}) reproduces (\ref{11}) in this limit.

Near the horizon, WKB approximation fails \cite{fro}. This can be
related to zero frequency modes of massive scalar fields appeared
in the Schwinger model \cite{moh}. So, to obtain the energy of
external charges near the horizon, we must use another
approximation.  Note that the solutions of (\ref{16}) in regions
near and far from the horizon must be matched asymptotically.

\section{ $q\bar{q}$ potential near horizon in the massive Schwinger model}

Near the horizon of a non-extremal black-hole, we can expand the
metric as
\begin{equation}\label{19}
f(r)=\kappa (r-h)+O(r-h)^2,
\end{equation}
where $\kappa:=df/dr$ is twice of the surface gravity. This metric
describes  Rindler space-time. In this region, $G(r,r')$ satisfies
\begin{equation}\label{20}
\left[-\partial_r\kappa(r-h)\partial_r+\mu^2+4\pi m\Sigma
\right]G(r,r')=\delta(r,r'),
\end{equation}
where $\Sigma=\mu\exp(\gamma)/\left(2\pi\right)$ \cite{ali} and
$\gamma$ is the Euler constant.  Two independent solutions of the
corresponding homogeneous equation are $I_{0}\left(2\bar{\mu}
\sqrt{(r-h)/\kappa}\right)$ and $K_{0}\left(2\bar{\mu}
\sqrt{(r-h)/\kappa}\right)$, where $I_0$ and $K_0$ are modified
Bessel functions. The well-defined Green function is then
\begin{equation}\label{21}
G(r,r')=\frac{2}{\kappa}K_{0}\left(2\bar{\mu}
\sqrt{\frac{r_{>}-h}{\kappa}}\right)I_{0}\left(2\bar{\mu}
\sqrt{\frac{r_{<}-h}{\kappa}}\right).
\end{equation}\label{22}
Defining $\mathcal{K}$ and $\mathcal{I}$ through
\begin{eqnarray}
 \frac{d\mathcal{K}}{dr}&=& K_{0}\left(2\bar{\mu}
 \sqrt{\frac{r-h}{\kappa}}\right),\nonumber \\
 \frac{d\mathcal{I}}{dr}&=& I_{0}\left(2\bar{\mu}
\sqrt{\frac{r-h}{\kappa}}\right),
\end{eqnarray}
then $\Pi$ in (\ref{15}) reduces to
\begin{eqnarray}\label{23}
\Pi &=&
-\frac{\mu^4}{\kappa}\Bigg\{\int_{r_a}^{r_b}W\left[\mathcal{I}(r),
\mathcal{K}(r)\right]dr
\nonumber \\
 &+&\left[\mathcal{I}(r_b)\mathcal{K}(r_b)+\mathcal{I}(r_a)\mathcal{K}(r_a)-
 2\mathcal{I}(r_a)\mathcal{K}(r_b)\right]\Bigg\},
\end{eqnarray}
where $W$ is the Wronskian. Considering the recurrence formulas
\begin{displaymath}\frac{d}{dx}\left[x^n
I_n(x)\right]=x^nI_{n-1}(x),\quad \frac{d}{dx}\left[x^n
K_n(x)\right]=-x^nK_{n-1}(x),\end{displaymath}
 one can show
\begin{eqnarray}\label{24}
\mathcal{K}(r)&=&-\frac{\sqrt{\kappa}}{\bar{\mu}}\left(r-h\right)^{\frac{1}{2}}K_1
\left(2\bar{\mu}\sqrt{\frac{r-h}{\kappa}}\right), \nonumber \\
\mathcal{I}(r)&=&\frac{\sqrt{\kappa}}{\bar{\mu}}\left(r-h\right)^{\frac{1}{2}}
I_1\left(2\bar{\mu} \sqrt{\frac{r-h}{\kappa}}\right).
\end{eqnarray}
Using  $W\left[K_1(x),I_1(x)\right]=1/x$, we arrive at
$W\left[\mathcal{I}(r),
\mathcal{K}(r)\right]=\kappa/(2{\bar{\mu}}^2)$.
 Therefore
\begin{eqnarray}\label{25}
E_{{\rm
elec.}}&=&\frac{e'^2}{2}\left(1-\left(\frac{\mu}{\bar{\mu}}\right)^2\right)
\left(r_b-r_a\right)
\nonumber \\
&-&\frac{e'^2\mu^2}{\kappa}\left[\mathcal{I}(r_b)\mathcal{K}(r_b)+\mathcal{I}
(r_a)\mathcal{K}(r_a)-2\mathcal{I}(r_a) \mathcal{K}(r_b)\right].
\end{eqnarray}
The first and the last terms describe the interaction of opposite
charges. While the first term corresponds to confinement aspects
of the model, the last term illustrates the screening effect. Note
that both the screening and confining phenomenon appear in the
same problem, similar to what happened in the flat case \cite{ar}.
For $m=0$ ($\mu=\bar{\mu}$), the confining term disappears. The
second and the third terms are the self-energies of the charges.
Writing eq.(\ref{25}) in the form (\ref{10}), we get
\begin{equation}\label{26}
\mathcal{G}(r,r')=\frac{1}{2}\left[1-\left(\frac{\mu}{\bar{\mu}}\right)^2\right
]\left(r_{>}-r_{<}\right)+\frac{2\mu^2}
{\kappa}\mathcal{I}(r_{<})\mathcal{K}(r_{>}).
\end{equation}
$\left(2\mu^2/\kappa\right)\mathcal{I}(r_{<})\mathcal{K}(r_{>})$
is the only term which contributes to the self-energy in
(\ref{25}). As this part of the Green function satisfies Dirichlet
boundary condition at the horizon, we find that $E_{{\rm
self.}}(r\rightarrow h)=0$. Far from the horizon, WKB
approximation is applicable and from (\ref{18}) we obtain $E_{{\rm
self.}}(r\gg h)=e'^2\mu^2/\left(4\bar{\mu}^3\right)$.

We can use the global method of Smith and Will \cite{will} to
determine the self-force. In a free falling system, the work done
by the force $F$ to displace slowly (such that the location of the
event horizon remains unchanged) the test charge by an
infinitesimal distance $\delta \bar{r}$ toward the horizon is
\begin{equation}\label{27}
\delta \bar{W}=-F\delta \bar{r}.
\end{equation}
The corresponding energy detected by an observer at asymptotic
infinity will be red-shifted
\begin{equation}\label{28}
\delta E=\sqrt{g_{tt}(r)}\delta \bar{W}.
\end{equation}
This change will be manifested by a change in the asymptotic mass
$\textbf{M}$ of the system, given by the total mass variation law
of Carter \cite{cart}
\begin{equation}\label{29}
\delta \textbf{M}=\delta \int_h^\infty T_t^tdr,
\end{equation}
where $T_t^t$ is the component of energy momentum tensor and is
the same as the effective Lagrangian in (\ref{7}). Therefore
$\delta \textbf{M}= \delta E_{{\rm self.}}$. Hence by transforming
locally the flat coordinates denoted by ${\bar r}$ to the
Schwarzschild ones, we obtain
\begin{equation}\label{30}
F=\frac{{\delta E}_{{\rm self.}}}{\delta r}.
\end{equation}

To derive (\ref{25}) we have assumed $|\phi(r)|\ll 1 $. Let us
check this assumption more carefully. Note that $\phi$ is the
solution of the equation (\ref{12}), obtained using eqs.(\ref{21})
and (\ref{22}) as:
\begin{equation}\label{31}
\phi(r)=\left\{
\begin{array}{ll} -\frac{2\mu e'}{\kappa}I_0\left(2\bar{\mu}
\sqrt{\frac{r-h}{\kappa}}\right)\left[\mathcal{K}(r_b
)-\mathcal{K}(r_a)\right] & \textrm{ $r<r_a<r_b,$}\\
 -\frac{2\mu e'}{\kappa}K_0\left(2\bar{\mu}
 \sqrt{\frac{r-h}{\kappa}}\right)\left[\mathcal{I}(r_b)
-\mathcal{I}(r_a)\right]& \textrm{$r>r_b>r_a,$}\\
 -\frac{2\mu e'}{\kappa}\Big[K_0\left(2\bar{\mu} \sqrt{\frac{r-h}{\kappa}}\right)
 \mathcal{I}(r)
 -I_0\left(2\bar{\mu}
 \sqrt{\frac{r-h}{\kappa}}\right)\mathcal{K}(r)+\nonumber \\
 + I_0\left(2\bar{\mu}
\sqrt{\frac{r-h}{\kappa}}\right)\mathcal{K}(r_b)
 -K_0\left(2\bar{\mu} \sqrt{\frac{r-h}{\kappa}}\right)
\mathcal{I}(r_a)\Big] & \textrm{$r_a<r<r_b.$}
\end{array} \right.
\end{equation}
The expressions appeared in $\phi(r)$ are in the forms
\begin{displaymath}
\mathcal{F}_1=\frac{2\mu e'}{\kappa}K_0\left(2\bar{\mu}
\sqrt{\frac{r_{>}-h}{\kappa}}\right)\mathcal{I}(r_{<}),\end{displaymath}
and \begin{displaymath}\mathcal{F}_2=\frac{2\mu
e'}{\kappa}I_0\left(2\bar{\mu}
\sqrt{\frac{r_{<}-h}{\kappa}}\right)\mathcal{K}(r_{>}).\end{displaymath}
Using the following  expansions near the horizon \cite{wolf},

\begin{eqnarray}\label{32}
K_0(x)&&\sim -\left[\gamma
+\ln\left(\frac{x}{2}\right)\right]\left(1+\frac{x^2}{4}\right)
+\frac{x^2}{4}+O(x^4),\nonumber \\
K_1(x)&&\sim \frac{1}{x}+\frac{x}{2}\left[\ln\left(
\frac{x}{2}\right)-\frac{\psi(1)+\psi(2)}{2}\right]+O(x^3),\nonumber \\
I_{0}(x)&& \sim 1+\frac{x^2}{4}+O(x^4),\nonumber \\
I_1(x)&& \sim \frac{x}{2}+O(x^3),
\end{eqnarray}
where $\psi$ is digamma function, one finds
\begin{eqnarray}\label{33}
\mathcal{F}_1&\sim& -\frac{2\mu e'(r_{<}-h)}{\kappa}\left[
\gamma+\ln\left(\bar{\mu}\sqrt{\frac{r_{>}-h}{\kappa}}\right)\right],
\nonumber \\
\mathcal{F}_2 && \sim \mu e'\Bigg\{\frac{1}{\bar{\mu}^2}
+\frac{r_{<}-h}{\kappa}\nonumber \\
&+&\frac{2(r_{>}-h)}{\kappa}
\left[\ln\left(\bar{\mu}\sqrt{\frac{r_{>}-h}{\kappa}} \right)
 -\frac{\psi(1)+\psi(2)}{2}\right]\Bigg \}.
 \end{eqnarray}
This shows that the assumption $|\phi|\ll 1$ is applicable near
the horizon, in which $r\rightarrow h$.

From the asymptotic expansions \cite{wolf}
\begin{eqnarray}\label{34}
I_{\nu}(z)&&\sim
\frac{1}{\sqrt{2\pi}}z^{\nu}\left(-z^2\right)^{-\frac{1}{4}(2\nu
+1)}\Bigg[e^{-i\left[\sqrt{-z^2}-\frac{1}{4}(2\nu
+1)\pi\right]}\left(1+O(z^{-1})\right)\nonumber \\
&+&{\rm c.c.}\Bigg]\nonumber \\
K_{\nu}(z)&&\sim
\sqrt{\frac{\pi}{2z}}e^{-z}\left[1+O(z^{-1})\right],
\end{eqnarray}
one can show that the eq.(\ref{25}) asymptotically reduces to
(\ref{18}), as expected. This is the requirement explained at the
end of section 2.

\section{Electrostatic self-energy in the second order approximation}

To study the effect of the curvature in higher order approximation
of self-energy, we must improve and refine the approximation
(\ref{19}) by including terms proportional to the scalar curvature
in the metric expansion near the horizon $r\simeq h$. We put
\begin{equation}\label{35}
f(r)=\kappa (r-h)+\frac{R}{2}(r-h)^2+O((r-h)^3),
\end{equation}
where $\kappa=f'(h)>0$ and $R=f''(h)$. This describes a space-time
with constant curvature $-R$ (locally AdS or dS, depending on
whether the sign of $-R$ is negative or positive, respectively).
In terms of the coordinate $u=x+b$, where
\begin{displaymath} b=\frac{\kappa}{R}, \qquad x=r-h,\end{displaymath}
the equation (\ref{16}) becomes
\begin{equation}\label{36}\left[
-\partial_u\frac{R}{2}\left(u^2-b^2\right)\partial_u+
\bar{\mu}^2\right]G\left(u, u'\right)=\delta\left(u, u'\right).
\end{equation}
In space-times with a constant positive curvature, $\Sigma$ is a
constant \cite{wipf}:
\begin{equation}\label{37} \Sigma=\frac{\mu
e^{\gamma}}{2\pi}\exp\left[\frac{1}{2}\left\{\ln\left(-\frac{R}{2\mu^2}\right)+
\psi(\frac{1}{2}+\alpha)+\psi(\frac{1}{2}-\alpha)\right\}\right],
\end{equation}
where $\alpha^2=1/4+2\mu^2/R$. We restrict ourselves to small
negative $R$, which we encounter in the next section where a
dilatonic black hole with a large mass is considered. For $|R|\ll
\mu^2 $,the chiral condensate becomes
\begin{equation}\label{39}
\Sigma=\frac{e^{\gamma}\mu} {2\pi}\exp\left[ \frac{R}{12\mu^2}
\right].
\end{equation}
Defining \begin{displaymath}z:=\frac{u}{b}, 0<z<1\qquad
;\tilde{\mu}^2:=\frac{-2\bar{\mu}^2}{R}\end{displaymath} the
homogeneous counterpart of the equation (\ref{36}) becomes
\begin{equation}\label{38}
\left[\left(1-z^2\right)\partial_z^2-2z\partial_z-\tilde{\mu}^2\right]G_h(z)=0.
\end{equation}

The real solutions of (\ref{38}) can be expressed in terms of
conical functions defined by \cite{nail}
\begin{eqnarray}\label{40}
p_{\tilde{\nu}}^n(z)&=&P_{\tilde{\nu}}^{-n}(z),\nonumber
\\
q_{\tilde{\nu}}^n(z)&=&
\frac{(-1)^n}{2}\left[Q^n_{\tilde{\nu}}(z)+Q^n_{-\tilde{\nu}-1}(z)\right]\nonumber \\
&=& -\frac{\pi}{2\sin(\pi\tilde{\nu})}P_{\tilde{\nu}}^n(-z).
\end{eqnarray}
$P_{\tilde{\nu}}^n(z)$ and $Q^n_{\tilde{\nu}}(z)$ are associated
Legendre functions and $\tilde{\nu}$ is defined through
$\tilde{\nu}(\tilde{\nu}+1)=-\tilde{\mu}^2$. Note that in the
small curvature limit $-2\bar{\mu}^2/R>\frac{1}{4}$ or
$\tilde{\mu}^2>\frac{1}{4}$, $\tilde{\nu}$ is a complex number.
Near the horizon, $z=1$, we have \cite{wolf}
\begin{eqnarray}\label{41}
\lim_{z\rightarrow 1}q_{\tilde{\nu}}(z)&=&-\frac{1}{2}
\ln\left(\frac{1-z}{2}\right)\left[1+O(1-z)\right]- \nonumber
\\ &&\frac{1}{2} \Big[\psi(1+\tilde{\nu})
+\psi(-\tilde{\nu})+2\gamma\Big]\left[1+O(1-z)\right],\nonumber\\
\lim_{z\rightarrow
1}p_{\tilde{\nu}}(z)&=&1-\frac{\tilde{\nu}(\tilde{\nu}+1)}
{2}(1-z)+O(1-z)^2.
\end{eqnarray}
Hence the real well behaved Green function is
\begin{equation}\label{42}
G(r,r')=\frac{2}{\kappa}p_{\tilde{\nu}}(z_{>})q_{\tilde{\nu}}(z_{<}).
\end{equation}
To arrive at this relation, we have used
$W\left[p_{\tilde{\nu}}^n(z),
q_{\tilde{\nu}}^n(z)\right]=1/\left(1-z^2\right)$ \cite{nail}.

Let us compare this result with one obtained in the previous
section. In terms of $\theta$ defined by $z=\cos{\theta}$,
(\ref{42}) becomes
\begin{equation}\label{43}
G(z,z')=\frac{2}{\kappa}p_{\tilde{\nu}}(\cos
\theta_{<})q_{\tilde{\nu}}(\cos \theta_{>}).
\end{equation}
One can make use of the relations \cite{nail}
\begin{eqnarray}\label{44}
\lim_{\theta\to 0}\tilde{\nu}^np_{\tilde{\nu}}^n(\cos
\theta)&=&i^nI_n(n\lambda), \nonumber
\\
\lim_{\theta\to 0}\tilde{\nu}^{-n}q_{\tilde{\nu}}^n(\cos
\theta)&=&i^{-n}K_n(n\lambda),
\end{eqnarray}
where $n\lambda /\left(\sin \theta \right)=\tilde{\mu}$, to obtain
\begin{equation}\label{45}
G(z,z')=\frac{2}{\kappa}I_0(\tilde{\mu}\theta_{<})K_0(\tilde{\mu}\theta_{>}),
\end{equation}
in the vicinity of the horizon, i.e. $ \theta\rightarrow 0$. In
this region we have also $x \approx -\theta^2 b/2$ which yields
$\theta \tilde{\mu}=2\sqrt{x{\bar{\mu}^2}/\kappa}$. Therefore in
the limit $x\rightarrow 0$, (\ref{42}) tends to (\ref{21}).

Putting back (\ref{42}) into (\ref{15}) we get
\begin{equation}\label{46}
\Pi=-\frac{\mu^4}{\kappa}W[\mathcal{P},
\mathcal{Q}](r_b-r_a)-\frac{\mu^4}{\kappa}\left[\mathcal{P}(r_a)\mathcal{Q}(r_a)
+\mathcal{P}(r_b)\mathcal{Q}(r_b)
-2\mathcal{P}(r_a)\mathcal{Q}(r_b)\right],
\end{equation}
where
\begin{eqnarray}\label{47}
\frac{d\mathcal{Q}(r)}{dr}&=&q_{\tilde{\nu}}\left(1+\frac{r-h}{b}\right),\nonumber
\\
\frac{d\mathcal{P}(r)}{dr}&=&p_{\tilde{\nu}}\left(1+\frac{r-h}{b}\right).
\end{eqnarray}
With the help of the recurrence formulas \cite{wolf}, \cite{bat}
\begin{eqnarray}\label{48}
\left(1-z^2\right)\frac{dP_{\tilde{\nu}}^{-1}(z)}{dz}&=&-\tilde{\nu}
zP_{\tilde{\nu}}^{-1}(z)+\left(\tilde{\nu}-1\right)P_{\tilde{\nu}-1}^{-1}(z),
\nonumber \\
\left(1-z^2\right)^{\frac{1}{2}}P_{\tilde{\nu}}(z)&=&z\left(\tilde{\nu}+1\right)
P_{\tilde{\nu}}^{-1}(z)-\left(\tilde{\nu}-1\right)
P_{\tilde{\nu}-1}^{-1}(z),
\end{eqnarray}
and
\begin{eqnarray}\label{49}
\left(1-z^2\right)\frac{dq_{\tilde{\nu}}^1(z)}{dz}&=&-\tilde{\nu}zq_{\tilde{\nu}}^1(z)
+\left( \tilde{\nu}+1\right)q_{\tilde{\nu}-1}^1(z),\nonumber
\\
-\tilde{\nu} \left(1-z^2\right)^{\frac{1}{2}}q_{\tilde{\nu}}(z)
&=& q_{\tilde{\nu}-1}^1(z) -zq_{\tilde{\nu}}^1(z),
\end{eqnarray}
one obtains
\begin{eqnarray}\label{50}
\mathcal{Q}(r)&=&-\frac{b}{\tilde{\nu}\left(\tilde{\nu}+1\right)}
\left(1-z^2\right)^{\frac{1}{2}}
q_{\tilde{\nu}}^1(z),\nonumber \\
\mathcal{P}(r)&=&-b\left(1-z^2\right)^{\frac{1}{2}}
p_{\tilde{\nu}}^1(z).
\end{eqnarray}
The first equation of (\ref{49}) is derived directly from the
first identity  of (\ref{48}). To verify the second equation, we
write it in terms of associated Legendre function $P$ as
\begin{equation}\label{51}
-P_{\tilde{\nu}-1}^1(z)+zP_{\tilde{\nu}}^1(z)=-\tilde{\nu}(1-z^2)^{\frac{1}{2}}
P_{\tilde{\nu}}(z).
\end{equation}
Using \cite{bat}
\begin{equation}\label{52}
zP_{\tilde{\nu}}^{\mu}(z)-P_{\tilde{\nu}+1}^\mu (z)=(\mu
+\tilde{\nu})(1-z^2)^{\frac{1}{2}}P_{\tilde{\nu}}^{\mu-1}(z),
\end{equation}
(\ref{51}) reduces to
\begin{equation}\label{53}
-P_{\tilde{\nu}-1}^1(z)+zP_{\tilde{\nu}}^1(z)=-\frac{\tilde{\nu}}{\tilde{\nu}+1}
\left[zP_{\tilde{\nu}}^1(z)-P_{\tilde{\nu}+1}^1(z)\right].
\end{equation}
But (\ref{53}) is nothing but the known equation: \cite{wolf}
\begin{equation}\label{54}
P_{\nu}^{\mu}(z)=\frac{2\nu+3}{\mu+\nu+1}zP_{\nu+1}^\mu
(z)+\frac{\mu-\nu-2}{\nu+\mu+1}P_{\nu+2}^\mu(z),
\end{equation}
when one takes $\nu=\tilde{\nu}-1$ and $\mu=1$.

 Using
$W[\mathcal{P},\mathcal{Q}]=b/\left[\tilde{\nu}\left(\tilde{\nu}+1\right)\right]$,
$E_{{\rm elec.}}$ from (\ref{14}) and (\ref{46}) becomes
\begin{eqnarray}\label{55}
E_{{\rm
elec.}}&=&\frac{e'^2}{2}\left(1-\left(\frac{\mu}{\bar{\mu}}\right)^2\right)
\left(r_b-r_a\right)
\nonumber \\
&-&\frac{e'^2\mu^2}{\kappa}\left[\mathcal{P}(r_b)\mathcal{Q}(r_b)+
\mathcal{P}(r_a)\mathcal{Q}(r_a)-2\mathcal{P}(r_a)
\mathcal{Q}(r_b)\right].
\end{eqnarray}
From
$\Gamma(\tilde{\nu})\Gamma(1-\tilde{\nu})=\pi/\left[\sin\left(\pi
\tilde{\nu}\right)\right]$, it follows that \cite{wolf}
\begin{eqnarray}\label{56}
\lim_{z\to 1}
p^{1}_{\tilde{\nu}}(z)&=&(1+z)^{-\frac{1}{2}}\left[(1-z)^{\frac{1}{2}}-
\frac{\tilde{\nu}(\tilde{\nu}+1)}
{2\Gamma(3)}(1-z)^{\frac{3}{2}}\right]+O(1-z)^{\frac{5}{2}},\nonumber  \\
\lim_{z\to 1}q^{1}_{\tilde{\nu}}(z)
&=&-\frac{\tilde{\nu}(\tilde{\nu}+1)}{2}(1+z)^{-\frac{1}{2}}
(1-z)^{\frac{1}{2}}\Bigg[\ln\big(\frac{1-z}{2}\big)-\psi(2)+\gamma
+ \nonumber \\ \psi(-\tilde{\nu})&+&\psi(\tilde{\nu}+1)
\Bigg]\Bigg[1+O(1-z)\Bigg]
+(1+z)^{-\frac{1}{2}}(1-z)^{-\frac{1}{2}}.
\end{eqnarray}
We have also \cite{wolf}
\begin{eqnarray}\label{57}
\psi(-\tilde{\nu})+\psi(\tilde{\nu}+1)&=&\psi(\frac{1}{2}
+\frac{i\lambda}{2})+\psi(\frac{1}{2}-\frac{i\lambda}{2})\nonumber \\
&=&2\Re\big(\psi(\frac{1}{2}+\frac{i\lambda}{2})\big),
\end{eqnarray}
where $\lambda=\sqrt{4\tilde{\mu}^2-1}$. For slowly varying
metrics $\lambda\gg 1$,  one can use the expansion $\psi(w)\sim
\ln w-\left(1/2w\right)-1/\left(12w^2\right)+O\big(w^{-4}\big)$,
\cite{wolf}, to arrive at
\begin{equation}\label{58}
\psi(-\tilde{\nu})+\psi(\tilde{\nu}+1)=\ln
\frac{-2\bar{\mu}^2}{R}+\frac{8\bar{\mu}^2R-R^2}
{48\bar{\mu}^4}+O(\frac{R^3}{\bar{\mu}^6}).
\end{equation}
Combining (\ref{55})-(\ref{58}) one finds
\begin{eqnarray}\label{59}
E_{{\rm self.}}(x)&=&\frac{e'^2}{2}\frac{\mu^2}{\bar{\mu}^2}
\Big\{x+\frac{\bar{\mu}^2}{\kappa}x^2[\ln
\frac{\bar{\mu}^2x}{\kappa}+ \gamma -\psi(2)+\frac{1}{2}
+\frac{R}{6\bar{\mu}^2}\nonumber
\\&-&\frac{R^2}{48\bar{\mu}^4} +O(R^3)]\Big\}+O(x^3).
\end{eqnarray}
Note that at $m=0$, where ${\bar \mu}=\mu$, the above expression
reduces to one obtained in \cite{moh}.
\section{Bekenstein entropy bound}
Our aim is now to use the relation (\ref{59}) to obtain the upper
entropy bound of a charged object. To do so, we allow the black
hole to carry a charge $q$ and assume $q$, $e'$ and $\mu$ to be
very small with respect to the black hole mass. We consider two
dimensional dilatonic charged black hole with mass $M$ and charge
$q$:
\begin{equation}\label{60}
ds^2=(1-2Me^{-r}+q^2e^{-2r})dt^2-
(1-2Me^{-r}+q^2e^{-2r})^{-1}dr^2,
\end{equation}
This metric emerges in the heterotic string theory as a solution
of the action
\begin{equation}\label{61}
S[g,\varphi,A]=\frac{1}{2} \int_{\mathcal{M}}
d^2x\sqrt{g}e^{\varphi}(R+(\nabla \varphi)^2- \frac{1}{2} F_{\mu
\nu}F^{\mu \nu})-\int_{\partial
\mathcal{M}}dx\sqrt{I}e^{\varphi}K,
\end{equation}
describing $2d$ gravity coupled to dilatonic field $\varphi$. $K$
is the extrinsic curvature  and $I$ is the induced metric on
$\partial \mathcal{M}$, where $\mathcal{M}$ is the surface under
study. The boundary term is added to make the variation procedure
self-consistent. The metric (\ref{60}) can support an
electrostatic test charge \cite{lin}. Thermodynamical quantities
for this system can be obtained using the Massieu function,
expressed in term of grand canonical partition function
corresponding to the action (\ref{61})\cite{fro3}. In $2d$, the
horizon surface of the black hole is a point and we cannot
consider the area. Nevertheless it might be useful to think about
the value of dilatonic field at the horizon, $\varphi_h$, as a
quantity playing the r\^ole of the logarithm of an effective area.
The event horizon of the black hole is located at
\begin{equation}\label{62}
h=\ln[M+(M^2-q^2)^{\frac{1}{2}}].
\end{equation}
For $q>M$ we have a naked singularity. The entropy of the system
is obtained as \cite{fro3}
\begin{eqnarray}\label{63}
S=2\pi e^{\varphi_h}=2\pi[M+(M^2-q^2)^\frac{1}{2}].
\end{eqnarray}
Comparing (\ref{60}) with (\ref{6}), gives
$f(r)=1-2Me^{-r}+q^2e^{-2r}$. Expanding $f(r)$ near $r=h$, results
$\kappa >0$ and $R<0$ for $q\ll M$. $\kappa$ and $R$ are defined
through eq.(\ref{35}).

 We now consider a charged object of rest mass $m$ and charge $e$
whose gravitational field is negligible on this background. This
object is slowly (adiabatically) descended toward the black hole.
This process causes no change in the horizon location and the
entropy of the black hole remains unchanged \cite{bek2}. To find
the change in black hole entropy caused by assimilation of the
object, one should evaluate the energy at the point of
capture,which is at a proper distance $l$ outside the horizon
\begin{equation}\label{64}
l=\int_0^a\frac{dy}{[\kappa y+\frac{Ry^2}{2}]^\frac{1}{2}}
=2\sqrt{\frac{a}{\kappa}}
-\frac{1}{6}R\big(\frac{a}{\kappa}\big)^\frac{3}{2}
+O\big(\frac{a}{\kappa}\big)^\frac{5}{2}.
\end{equation}
$y$ is the coordinate distance from the horizon and  $a$ is the
position of the center of mass of the object. In fact $l$ is the
proper radius of the object at the point of capture. For small $l$
we get $\sqrt{a/\kappa}=l/2+\left(R/96\right)l^3+O(l^5)$. When the
object is assimilated, its charge modifies the hole's charge to
$q+e'$, and its total energy, which we denote by $E_T$, augments
the hole's mass from $M$ to $M'=M+E_T$.

The energy of the object, $E_T$, is constituted of the energy of
the body's mass, $m'$, shifted by the gravitational field,
$E_{m'}\simeq m'\kappa l/2+O(l^3)$, and the electrostatic
self-energy $E_{{\rm self.}}$ (\ref{59}) which in terms of $l$ is:
\begin{eqnarray}\label{65}
E_{{\rm
self.}}&=&\frac{e'^2}{2}\frac{\mu^2}{\bar{\mu}^2}\Bigg\{\frac{\kappa
l^2}{4}+\frac{\kappa Rl^4}{96} +\frac{\bar{\mu}^2\kappa
l^4}{16}\bigg[ 2\ln\left(\bar{\mu}\left(
\frac{l}{2}+\frac{Rl^3}{96}\right)\right)+\gamma-\psi(2)
\nonumber \\
&+&\frac{1}{2}+\frac{R}{6\bar{\mu}^2}-\frac{R^2}{48\bar{\mu}^4}+O(R^3)
\bigg]\Bigg\}+ O(l^6).
\end{eqnarray}
The final entropy of the black hole is
\begin{equation}\label{66}
S_f=2\pi\left(2M+2E_T-\frac{q^2+e'^2+2qe'}{2M}+O(M^{-2})\right).
\end{equation}
In the massive Schwinger model, the fermionic parameters $\mu$ and
$m$ are appeared in the final black hole entropy through the
self-energy $E_T$. In contrast, in the massless Schwinger model,
where $\bar{\mu}=\mu$, any information about fermionic field is
lost (up to the order $l^4$) . This result agrees with \cite{moh}.

Assuming the validity of the generalized second law of
thermodynamics \cite{ho}
\begin{equation}\label{67}
S_f\geq(S_i + S),
\end{equation}
where $S_{f(i)}$ is the black hole entropy in final (initial)
state and $S$ is the object entropy, we obtain
\begin{equation}\label{68}
S\leq 4\pi E_T -\frac{\pi e'^2}{M}-\frac{2\pi e'q}{M}.
\end{equation}

When $M$ is large with respect to $e$, $q$, $m'$, and $e'$, we can
expand the surface gravity \begin{equation} \kappa
=2\sqrt{M^2-q^2}/\left(M+\sqrt{M^2-q^2}\right), \end{equation} as
$\kappa\simeq 1-q^2/\left(4M^2\right)+O(M^{-4})$. In this limit
the leading terms in the inequality (\ref{68}) are independent of
black hole's parameters
\begin{equation}\label{69}
S\leq 2\pi m'l+\frac{\pi
e'^2}{2}\frac{\mu^2}{\bar{\mu}^2}l^2+O(\frac{1}{M}).
\end{equation}
This shows that the electrostatic self-energy modifies the
Bekenstein upper bound.

In the massless Schwinger model, $\mu={\bar \mu}$, the
eq.(\ref{69}) becomes
\begin{equation}\label{70} S\leq 2\pi m'l+\frac{\pi
e'^2}{2}l^2+O(\frac{1}{M}),
\end{equation}
which indicates the absence of the fermionic information in the
upper entropy bound of the object. This agrees to the fact that
near the horizon, massive gauge fields act like conformal massless
fields. Besides the r\^{o}le of vacuum polarization, a main
difference with respect to the result obtained in $QED_4$,
\cite{zala}, is the sign and the order of terms in right hand side
of (\ref{69}), in other words the bound is not tightened here. For
$e'=0$ we obtain the well known result \cite{mig}
\begin{equation}\label{71}
S\leq 2\pi m'l.
\end{equation}

\end{document}